\documentclass[conference]{IEEEtran}

\usepackage{cite}
\usepackage{graphicx}
\usepackage{array}
\usepackage{tikz}
\usepackage{tikz-3dplot}
\usetikzlibrary{shapes.geometric}
\usetikzlibrary{decorations.pathreplacing}
\usetikzlibrary{positioning}
\usetikzlibrary{patterns}
\usepackage{pgfplots}
\usepgfplotslibrary{fillbetween}
\usepackage{pstool}  
\usepackage{mathtools}
\usepackage[linesnumbered,algoruled,boxed,lined]{algorithm2e}
\usepackage[absolute]{textpos}
\usepackage{threeparttable}
\usepackage[strict]{changepage}
\usepackage[keeplastbox]{flushend} 
\usepackage{amssymb}
\usepackage{comment}
\usepackage{bbm}
\usepackage{hhline}
\usepackage{lipsum} 
\usepackage{enumitem}
\usepackage{soul}
\usepackage{xcolor}

\usepackage{amsthm}

\ifCLASSINFOpdf
\else
\fi
\usepackage{amsmath}
\ifCLASSOPTIONcompsoc
  \usepackage[caption=false,font=normalsize,labelfont=sf,textfont=sf]{subfig}
\else
  \usepackage[caption=false,font=footnotesize]{subfig}
\fi
\newcommand{\dif}{\mbox{d}}

\pgfplotsset{compat=1.14} 

\begin{document}
%
\title{Intermodulation Interference Detection in 6G Networks: A Machine Learning Approach}
%


\author{\IEEEauthorblockN{Faris B. Mismar}
\IEEEauthorblockA{Nokia Bell Labs Consulting, Murray Hill, NJ, 07974}
}

\maketitle
\begin{abstract}
This paper demonstrates the use of machine learning to detect the presence of intermodulation interference across several wireless carriers.  We show a salient characteristic of intermodulation interference and propose a machine learning based algorithm that detects the presence of intermodulation interference through the use of supervised learning. This algorithm can use the radio access network intelligent controller or the sixth generation of wireless communication (6G) edge node as a means of computation. Our proposed algorithm runs in linear time in the number of resource blocks, making it a suitable radio resource management application in 6G.

\end{abstract}


%
\IEEEpeerreviewmaketitle

\section{Introduction} 
%
%
%
%



 
As more frequency bands are introduced to the fifth generation of wireless networks (5G) and beyond (B5G), the combinations of the carriers used on the base stations increase. Combined with sources of intermodulation interference, which can be difficult to spot in the coverage area of a cell, the detection of intermodulation interference becomes ever important.  The mitigation of intermodulation interference can help wireless operators improve the performance of the cell.

Intermodulation interference can be either active or passive based on the existence of active elements in the radio circuit. Active intermodulation interference can be avoided by not driving the amplifiers past their linear region.  What remains is the passive intermodulation interference.  Passive intermodulation can happen when deploying multiple frequency bands on a single network element (e.g., an antenna) to save physical space.  Sources of passive intermodulation interference are either \textit{internal} such as corroded junctions, duplexers, or cables that are contaminated with metallic particles, or \textit{external} such as rusty bolts or fences. We will use the term \textit{intermodulation interference} to refer to the passive type in this paper.


With the development of wireless communications systems beyond 5G, network and computing convergence platforms in the radio access network (RAN) such as the sixth generation of wireless communication (6G) ``multi-access edge computing'' are expected to be one of the enablers for machine learning applications that can operate in near real-time.  {Another example of these enablers is also present in the Open-RAN standards such as the RAN intelligent controller \cite{o-ran}.}  While the topic of intermodulation interference detection is not new, it is the use of machine learning based algorithms that can detect intermodulation near real-time and without test signals or noise simulators that makes this topic not only suitable for present 5G, but also---and through the use of edge computing---suitable for successive network evolutions.

Prior work related to detection of intermodulation interference was studied in \cite{8706428}. A signal modeling method was proposed using block computations methods on third-order intermodulation products.  This effectively has two limitations: 1) it focuses only on detection of third-order intermodulation products and 2) its processing delay per computational step is dependent on the sample (or the ``block'') size.  Our proposed method is not limited to third-order intermodulation products and has linear run-time complexity. In \cite{9500204}, the use of anomaly detection as a technique to detect performance abnormalities such as excessive uplink power was proposed.  However, while the excessive uplink power anomaly could be a sign of intermodulation interference in the uplink frequency band, it could also be due to other impairments not related to intermodulation interference such as narrow-band interference.  Industry standards such as \cite{3gpp38101} propose solutions that generate channel noise as a means to detect intermodulation interference.  However, this brings \textit{further} load and interference to a BS already suffering from impairments.  

In this paper, we provide an answer to whether a reliable non-surgical method for intermodulation interference detection can be derived. 

This paper makes two specific contributions:
\begin{enumerate}
\item Motivate why the intermodulation interference power spectral density in the presence of Fast Fourier Transform (FFT) is sloped and thus can be approximated by a line.
\item Use a machine learning algorithm based on simple linear regression that can detect intermodulation interference in real-time and \textit{without} any surgical activities (e.g., injection of test tones or channel noise simulators).
\end{enumerate}%

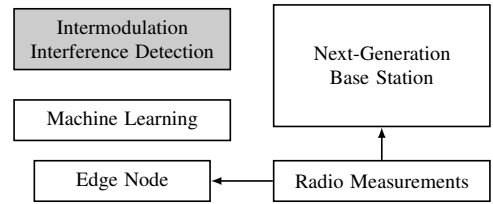
\begin{figure}[!t]
\centering
\resizebox{0.35\textwidth}{!}{\begin{tikzpicture}[style=thick, node distance=2cm, scale=1, >=latex]
    \node [coordinate, name=input] {};
	\node [rectangle, draw, right of=input, 
		text width=10em, text centered, minimum height=6em] (BS) {Next-Generation Base Station};
    
    \node [rectangle, draw, below of=BS, node distance=2cm,
 		text width=10em, text centered, minimum height=2em] (RIP) {Radio Measurements};	
    
    \node [rectangle, draw, left of=RIP, node distance=3.5cm, xshift=-1cm,
		text width=8em, text centered, minimum height=2em] (RIC) {Edge Node};
    \node [rectangle, draw, above of=RIC, node distance=3em,
		text width=10em, text centered, minimum height=2em] (AI) {Machine Learning};

    \node [rectangle, fill=gray!40, draw, above of=AI, node distance=4em, 
        text width=10em, text centered, minimum height=3em] (unsup) {Intermodulation Interference Detection};

    \draw [draw,->] (RIP.180) -- node {} (RIC.0);
    \draw [draw,<-] (BS.270) -- node {} (RIP.90);

\end{tikzpicture}%
}%
\caption{Next-generation base station using machine learning in an edge node on radio measurements to detect the existence of intermodulation interference.}
\label{fig:overall}
\end{figure}

\vspace*{-0.5em}
\section{System Model}

We consider a system composed of a single base station (BS) with multiple transmit (i.e., downlink) and receive (i.e., uplink) frequencies each of a bandwidth $B$.  Intermodulation interference sources and user equipment devices (UEs) are independently scattered in the service area of the BS.  The system uses frequency division duplex (FDD) mode of operation, even though time division duplex (TDD) is also possible jointly with an FDD deployment or when different uplink-downlink frame configurations are elected for different frequency bands.

The system uses orthogonal frequency division multiplexing (OFDM) in both uplink and downlink\footnote{We use OFDM as an umbrella term that may include proposed 6G variants (e.g., the Discrete Fourier Transform spread OFDM waveform in 5G).}.  For simplicity of notation in the representation of the OFDM waveform in the time domain, we choose the {$k$-th OFDM subcarrier} at the {time instant $n$} using quadrature amplitude modulation and center frequency $f_c$ and {write:%
\begin{equation}
    x_{c,k}[n] =  \text{Re}\left ( g_B(t - nT) A_{c,k} e^{j 2\pi k \Delta f (t - nT)} e^{j 2\pi f_c t}  \right )
    \label{eq:ofdm}
\end{equation}%
\noindent where $A_{c,k}$} is a complex signal of the inphase and quadrature components, $\Delta f$ is the OFDM subcarrier spacing, {$T$ is the subcarrier duration}, and $g_B(t)$ is a \textit{windowed} sinc pulse shaping function with bandwidth $B$.  There are $N_\text{SC}$ subcarriers in one physical resource block (PRB).  The number of subcarriers $N_\text{SC} = \lfloor B / \Delta f\rfloor$.  A PRB is the smallest non-overlapping frequency resource allocated in the fourth generation of wireless communication and its present evolutions \cite{3gpp38300}.  The number of PRBs available for allocation ($N_\text{PRB} > 0$) depends on the bandwidth $B$.

To potentially have intermodulation interference on the uplink or receive frequency band $B^\prime$, at least two subcarriers (one from each transmit frequency) have to be combined at a non-linearity (e.g., intermodulation interference sources) such that their product falls in band with the uplink frequency band. We represent this non-linearity as an $m$-th order polynomial $q\colon x\mapsto \sum_{k=1}^m a_kx^k$, with $a_k\neq 0$.  

\section{Intermodulation Interference}\label{sec:intermodulation}
Intermodulation interference may happen in any direction (i.e., uplink or downlink).  However, it is particularly interesting in the uplink since UEs are power-limited (i.e., compared to BSs).  This can make the uplink target signal-to-interference-plus-noise ratio more difficult to achieve causing access blockage and increased bit error rate.  Intermodulation interference can happen due to two or multiple center frequencies and can be of any order $m \ge 2$.

For simplicity of math, we study the intermodulation interference due to two center frequencies, each being a subcarrier belonging to a distinct PRB, and up to the \textit{third} intermodulation interference order.  Let these two frequencies be $f_1$ and $f_2$ and the subcarriers $x_{1,k}$ and $x_{2,\ell}$ as follows from \eqref{eq:ofdm}. Let us call the BS the transmit frequencies of which cause the intermodulation interference an ``offender.''

\textbf{Non-linearity:} When non-linearity is introduced between the voltage and current components of a signal, harmonic frequencies and linear combinations of them are generated.  For our signals $x_{1,k}[n]$ and $x_{2,\ell}[n]$, we consider an input signal
\begin{equation}
    \begin{aligned}
        x_{(k,\ell)}[n] &= x_{1,k}[n] + x_{2,\ell}[n].
    \end{aligned}\label{eq:input}
\end{equation}
This input signal, when passing through a non-linearity with $m = 3$ creates an output signal $y_{(k,\ell)}[n]\coloneqq q(x_{(k,\ell)}[n])$:
\begin{equation}
    \begin{aligned}
        y_{(k,\ell)} &\coloneqq q(x_{1,k} + x_{2,\ell})  \\
            &= a_1(x_{1,k} + x_{2,\ell})  + a_2(x_{1,k} + x_{2,\ell})^2  \\
            &\qquad + a_3(x_{1,k} + x_{2,\ell})^3.\\
    \end{aligned}\label{eq:output}
\end{equation}
It is straightforward to substitute $x_{1,k}$ and $x_{2,\ell}$ in all the terms above with their respective OFDM waveform representation in \eqref{eq:ofdm}.  Examining the terms of $y_{(k,\ell)}$, there are first-order (or linear) terms which contain the \textit{fundamental} center frequencies $f_1$ and $f_2$. However, it is the quadratic and cubic terms that are of particular interest.  Expanding these higher-order terms and using trigonometric identities, we obtain OFDM subcarriers at the following center frequencies: $\{f_1, f_2, 2f_1, 2f_2, f_1 \pm f_2, 3f_1, 3f_2, 2f_1 \pm f_2, 2f_2 \pm f_1\}$.  Excluding the fundamental frequencies, we are left with second- and third-order harmonics.  These could interfere with other receive bands in the vicinity.%
\begin{figure}[!t]
\centering
\resizebox{0.35\textwidth}{!}{\begin{tikzpicture}[style=thick, node distance=0cm, >=latex]

\draw[-latex] (-4.5,0) -- (3.5,0) node[right] {$f$};
\draw[-latex] (-4.5,-0.5) -- (-4.5,2.5) node[above] {$P$};
    
 \foreach \x in {-3,-2,...,2.5}
        \draw (\x cm,0pt) -- (\x cm,-3pt)
            node[anchor=north,xshift=-0.15cm] {};
    \foreach \y/\ytext in {0.5/P_\text{IM5},1/P_\text{IM3},2/P_c}
        \draw (-4.5cm+0pt,\y cm) -- (-4.5cm-3pt,\y cm) node[anchor=east] {$\ytext$};

\draw[-latex,dashed] (-1,0) -- (-1,2) node[right] {$f_1$};
\draw[-latex,dashed] (0,0) -- (0,2) node[right] {$f_2$};

\draw[-latex] (-2,0) -- (-2,1) node[right] {$f_\text{IM3}$};
\draw[-latex] (1,0) -- (1,1) node[right] {$f_\text{IM3}$};
\draw[-latex] (-3,0) -- (-3,0.5) node[right] {$f_\text{IM5}$};
\draw[-latex] (2,0) -- (2,0.5) node[right] {$f_\text{IM5}$};




\end{tikzpicture}%
}%
\caption{OFDM tones (dashed) and their respective odd-order intermodulation products (solid).} 
\label{fig:pim}
\end{figure}
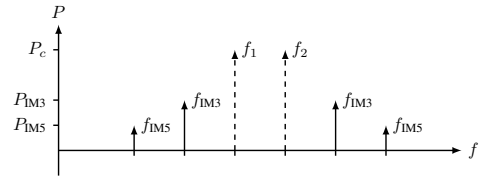%

\begin{figure}[!t]
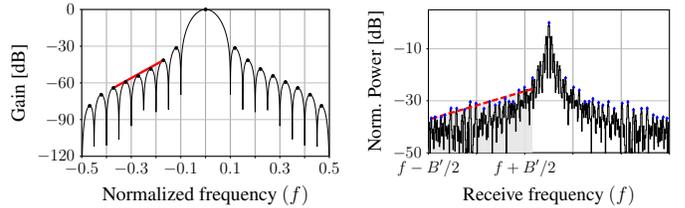

\captionsetup[subfigure]{labelformat=empty}
\begin{adjustwidth}{-1em}{0cm}
\centering
\subfloat[]{
\resizebox{0.25\textwidth}{!}{\vspace*{-10em} \input{figures/hanning_fft.tikz}}\label{fig:time}}
\hfil
\subfloat[]{
\resizebox{0.23\textwidth}{!}{\vspace*{-10em} \input{figures/sp_leakage.tikz}}\label{fig:freq}}
\end{adjustwidth}
\vspace*{-1em}\caption{Windowing spectral plots: (a) gain vs frequency plot of a window function (black) showing the side lobes (dotted peaks) and a line (red) the slope of which is equal to the rolloff rate (b) spectral leakage impact on intermodulation interference  (black) in a receive frequency band $B^\prime$ (shaded) with side lobe peaks (blue dots) and a fitted line (dashed red) of the windowed intermodulation products.}
\label{fig:spectral}
\end{figure}%

\textbf{Intermodulation order:} The sum of the absolute values of the frequency multipliers define the intermodulation order.  So, in the case of $\{2f_1\pm f_2, 2f_2 \pm f_1\}$, we have third-order intermodulation products ($f_\text{IM3}$). Let us call the portion of the output signal that has these third-order products $y_\text{IM3}$.  A similar computation can be made for a fifth-order non-linearity.  These are $\{3f_1 \pm 2f_2, 3f_2 \pm 2f_1\}$ and are the fifth-order intermodulation products ($f_\text{IM5}$). Similarly, the portion of the output signal that has those fifth-order products is $y_\text{IM5}$.  Fig.~\ref{fig:pim} shows the third- and fifth-order intermodulation subtractive products.  These are the combinations (or ``products'') involving a frequency difference and thus typically fall in-band of the receive PRBs.  The third- and fifth-order products have a bandwidth of $3B$ and $5B$.

\textbf{Intermodulation and duplex mode:} In FDD, the duplex gap (i.e., the difference between the downlink and uplink frequencies) dictates whether the subtractive products fall in the uplink frequency band.  However, in TDD, the transmit and receive frequency bands are the same.  Despite that, an odd-order intermodulation product (subtractive or additive) can fall in-band another TDD (or even FDD) receive frequency.  In fact, with several frequency bands allocated for transmission and reception spanning sub-6 GHz waves, millimeter waves, and terahertz waves, it is possible for either even-ordered or additive intermodulation products (e.g., $f_1 + f_2$ or $2f_1 + 2f_2$) to fall in-band one or more receive PRBs regardless of the duplex mode.

\textbf{Intermodulation characteristics:} We study the bandwidth, amplitude, and frequency difference between the intermodulation product and the fundamental frequency. Let us we start with a relaxed version of the signal \eqref{eq:ofdm} where we set $A_{c,k} = 1$ and compute the magnitude of the analytic part of the Fourier transform of $x_{c,k}$  denoted as $X_{c,k}(f)$:%
\begin{equation}
    \begin{aligned}
        \vert X_{c,k}(f) \vert = \frac{1}{\vert \tilde B\vert } \bigg  [ \text{rect} & \left (\frac{f - f_c - k\Delta f)}{B} \right )  \bigg ], 
    \end{aligned}
\end{equation}%
which is a rectangular pulse, with a base length (or \textit{support}) of $B$. Here, $\tilde B > 1$ is a collective term of coefficients. For the second-order intermodulation product ($p = 2$) in $y_{k,\ell}[t]$, this product in the time domain is a convolution between two $\text{rect}(\cdot)$ functions in the frequency domain.  From this we make the following three observations:
\begin{enumerate}
    \item Since each function has a support of $B$, their convolution $Y_{k,\ell}(f)$ is a triangle function with a support (or \textit{bandwidth}) of $2B$. The convolution of this triangle with another $\text{rect}(\cdot)$ for the third-order product creates a function with a support of $3B$.
    \item  The amplitude of the convolved signal reduces as $p$ increases. We find that the transmit power of the OFDM subcarrier $P_1 = 1/(T\vert \tilde B\vert ^2)$.  For the case of $p=2$, we compute the power of the second-order intermodulation product as {$P_2 \simeq 1/(T\vert \tilde B \vert^3)$}, which is smaller than $P_1$.
    \item The frequency difference of the intermodulation product center frequency to the fundamental center frequencies increase as $p$ increases.  For the case of $p = 3$, the difference between $f_\text{IM3} = 2f_1 - f_2$ and the fundamental frequency $f_1$ is $f_1 - f_2$.  However, for $p = 5$, $f_\text{IM5} = 3f_1 - 2f_2$ this difference from $f_1$ becomes $2(f_1 - f_2)$.
\end{enumerate}%
From these three findings, it follows that depending on the intermodulation product order $p$, the bandwidth $B$, and the fundamental frequencies, the intermodulation products can add up and fill up to a bandwidth of $pB$.  In OFDM-based systems, where FFT is used for modulation, frequency content due to a windowing pulse-shaping function brings forward the concept of ``spectral leakage.'' \cite{8736965}.  Windowing and spectral leakage are shown in Fig.~\ref{fig:spectral} where interference is spread across the entire band due to the frequency-domain convolution of the received signal with this windowing function at demodulation. This happens when a non-integer number of signal periods is sent to the FFT.  System designs account for this constraint, which leaves interference as a cause of this leakage. Due to this and the three findings, the ability to detect the presence of intermodulation interference in such systems becomes as simple as detecting a sloped line in the power spectral density.

\begin{figure}[!t]
\centering
\resizebox{0.48\textwidth}{!}{\begin{tikzpicture}[style=thick, node distance=0cm, >=latex]
\tikzstyle{every node}=[font=\footnotesize]
\draw (-5,0) -- (5,0) node[right] {PRB};


\draw[draw=black] (-4.5,0) rectangle (-2.5,0.5) node[midway] {Control Plane};
\draw[draw=black] (-2.5,0) rectangle (2.5,0.5) node[midway] {User Plane};
\draw[draw=black] (2.5,0) rectangle (4.5,0.5) node[midway] {Control Plane};

\draw (-4.5 cm,0pt) -- (-4.5 cm,-3pt) node[anchor=north,xshift=-0.15cm] {$0$};
\draw (-2.5 cm,0pt) -- (-2.5 cm,-3pt) node[anchor=north,xshift=-0.15cm] {$N_\text{PRB}^{(c)}/2 - 1$};
\draw (2.5 cm,0pt) -- (2.5 cm,-3pt) node[anchor=north,xshift=-0.15cm] {$N_\text{PRB} - N_\text{PRB}^{(c)}/2 - 1$};
\draw (4.5 cm,0pt) -- (4.5 cm,-3pt) node[anchor=north,xshift=0.15cm] {$N_\text{PRB} - 1$};

\end{tikzpicture}%
}%
\caption{Allocation of the control- and user-plane channels to the uplink PRBs.}
\label{fig:prbs}
\end{figure}
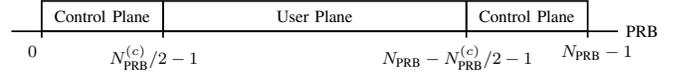%

\textbf{Interference measures:} If we let $P_c$ be the received power that correspond to one of the two fundamental frequencies in \eqref{eq:output} and let the white Gaussian noise power spectral density be equal to $N_0$ W/Hz, then we can write the received total power (RTP) in a given bandwidth as:%
\begin{equation} \label{eq:rtp}
    \textsf{RTP} \coloneqq P_c + N_0B + P_I +  \frac{1}{T}\sum_{p=2}^m \int_{pB} \vert Y_{\text{IM}p}(f) \vert ^2\, \dif f,
\end{equation}%
where {$P_I$ is a term representing interference not due to intermodulation}, $P_{\text{IM}p}$ is the in-band $p$-th order intermodulation product interference power, and $Y_{\text{IM}p}$ is the Fourier transform of $y_{\text{IM}p}$, which was defined for a given $p$ earlier in this section.  In the case of multi-input multi-output (MIMO) systems, RTP is computed per MIMO receive branch.  These receive branches are independent and have separate hardware.   The relationship between $P_c$ and $P_\text{TX}$ is governed by the path loss as we show here later.  Also, if we let $Y^{(\mathcal{P})}_{\text{IM}p}(f)$ be the output signal computed over the OFDM subcarriers in the PRB $\mathcal{P} \coloneqq \{\mathcal{P}_i\}_{i=0}^{N_\text{PRB} - 1}$, then with a little notation abuse we can write the RTP \eqref{eq:rtp} for the $j$-th received branch as:%
\begin{equation}\label{eq:rtp_alt}
\textsf{RTP}_j\coloneqq P_{c,j} + N_0B + P_{I, j} + \frac{1}{T}\sum_{i=0}^{N_\text{PRB} - 1} \sum_{p=1}^{m} \int_{pB} \vert Y^{(\mathcal{P}_i)}_{\text{IM}p,j}(f)\vert ^2 \,\dif f\!.
\end{equation}%

\textbf{Internal or external:} Receive branches of a given antenna typically have identical hardware configuration and hence report similar RTP values.  However, if RTP values differ and intermodulation interference is detected, then this can be due to an \textit{internal} source of intermodulation interference.  Also, if intermodulation interference is detected but the RTP values per receive branch are similar, then the source can be \textit{external}.

It also follows from \eqref{eq:rtp_alt} that one way to mitigate intermodulation interference is to reduce the transmitted OFDM tone power of the \textit{offender}, which is directly related to the BS transmit power ($P_c \coloneqq P_\text{TX} - L = P_\text{BS} - 10\log N_\text{PRB} - 10\log N_\text{SC} - L$), where $P_\text{BS}$ is the BS transmit power in dBm and $L$ is the path loss \cite{8665922}.  This power reduction also reduces the BS coverage radius.

\indent At this stage in the development, it is important to mention that the standards also allocate certain PRBs for control-plane channels, while others are used for the user-plane channels.  Control channels are allocated the PRBs at the edges of the bandwidth.  Let there be $N_\text{PRB}^{(c)}$ PRBs allocated for control-plane channel and $N_\text{PRB}^{(u)}$ allocated for the user-plane channel such that $N_\text{PRB}^{(c)} + N_\text{PRB}^{(u)} = N_\text{PRB}$, an example of which is shown in Fig.~\ref{fig:prbs}. Hence $\mathcal{P} = \mathcal{P}^{(c)} \cup \mathcal{P}^{(u)}$, written as a union of the sets of control- and user-plane PRBs. 

An important quantity is the received interference power (RIP), which is measured in Watts per PRB---as defined in \cite{3gpp36214}---since PRBs are the smallest frequency resource allocated. For the $r$-th PRB where $r \in \{0,1,\ldots,N_\text{PRB}-1\}$, we write:%
\begin{equation}\label{eq:rip}
    \begin{aligned}
        \textsf{RIP}(r) \coloneqq \sum_{p=1}^m 
        \int_{pB^{(r)}} & N_0\,\text{rect} \left ( \frac{f-f_{\text{IM}p}}{pB^{(r)}} \right )\dif f +  P_I^{(r)} + \\
        & \frac{1}{T} \sum_{p=2}^m  \int_{pB^{(r)}} \vert Y_{\text{IM}p}(f)\vert^2 \dif f, \\ 
    \end{aligned}
\end{equation}
where $B^{(r)}$ is the bandwidth of the $r$-th PRB in $\mathcal{P}$ and $P_I^{(r)}$ is a term representing interference not due to intermodulation as measured by the $r$-th PRB.  Plotting the RIP against the PRBs in $\mathcal{P}$ is a plot of noise plus interference power vs frequency and is therefore a power spectral density plot.

Owed to the selection of a windowing function and its side lobe rolloff rate \cite{ni_fft}, the power spectral density of spectral leakage (in dBm units) in the presence of interference can be approximated as a sloped line with a slope equal to the rolloff rate.  Otherwise, RIP---as in \eqref{eq:rip}---represents the power of white Gaussian noise, which is spectrally flat.  Therefore, a sloped line is a salient feature of intermodulation interference in an OFDM-based system; it necessary to detect the presence of intermodulation interference due to an offender, but it is not sufficient.  However, since wireless spectrum is often distinctly associated with a wireless operator, we discount the possibility of a spurious spectrally sloped in-band transmission, making this condition \textit{practically} sufficient to detect the presence of intermodulation interference.

Implementation dependent, the control-plane PRBs can have \textit{equal} power levels on both sides of the spectrum---unlike the user-plane PRBs.  This is because the control plane bandwidth is allocated per UE as one PRB on each side of the channel \cite{3gpp36211}, providing the UEs with robustness against frequency-selective fading.


\section{Intermodulation Interference Detection}\label{sec:detection}


\subsection{Current methods of detection}
\textbf{Introduction of test tones:} International standards for PIM testing \cite{iec,3gpp37808} specify two continuous wave tones to be injected at a fixed transmit power in the BS feeders to the antennas while the third-order intermodulation product {(IM3)} is monitored over a high-sensitivity receiver (e.g., with a spectrum analyzer). If IM3 products are observed (e.g., Fig.~\ref{fig:pim}), then intermodulation interference is {deemed detected in this BS}.  This approach has limitations: 1) it cannot be carried out in real time and 2) it would require the disconnection of the BS, hence the disruption of service for during the testing.

\textbf{OFDM channel noise generator:} The OFDM channel noise generator (OCNG) is specified in the industry standards for 5G \cite{3gpp38101}. It introduces channel noise as simulated load on the downlink frequency bands for both FDD and TDD.  Once enabled, the RTP of uplink frequency bands is monitored for an anomaly (e.g., abrupt rise) at the time the OCNG was activated.  Anomaly that follows the OCNG activation/deactivation patterns means that intermodulation interference is detected.  Testing with OCNG often requires loading the BS to the maximum (i.e., on all PRBs), which would generate interference and impact the channel quality and capacity on the PRBs belonging to the neighboring BSs.%

\begin{algorithm}[!t]
\small
    \caption{Intermodulation Interference Detection}
    \label{alg:algorithm}
    \DontPrintSemicolon
    \KwIn{RIP data $\mathbf{R}\in\mathbb{R}^{\vert\mathcal{T}\vert\times N_\text{PRB}^{(u)}}$ for a BS with $N_\text{PRB}$ PRBs.}
    \KwOut{Whether intermodulation interference is detected in the BS for the time of interest $t\in\mathcal{T}$}
    Let $\mathcal{P}^{(u)} \coloneqq \{\text{PRB}_i\colon  N_\text{PRB}^{(c)}/2 \le i \le N_\text{PRB} - N_\text{PRB}^{(c)}/2 - 1\}$\;
    Initialize $N_{\text{IM},t} \coloneqq \text{False}$\;
    $\mathbf{r} \coloneqq [\mathbf{R}]_{t,\mathcal{P}^{(u)}}$ \;
    $(R^2, \beta_1) \coloneqq \textbf{LinearRegression}(\mathcal{P}^{(u)}, \mathbf{r}_t)$\;
    \If {($R^2 > \varepsilon_\text{\rm linear}$) {\bf and} ($\vert \beta_1 \vert > \varepsilon_\text{\rm slope}$)} {
        $N_{\text{IM},t} \coloneqq \text{True}$
    }    
    \Return $N_{\text{IM},t}$\;
\end{algorithm}%

\begin{figure*}[!t]
\begin{adjustwidth}{-1em}{0cm}
\centering
\subfloat[]{
\resizebox{0.24\textwidth}{!}{\vspace*{-10em} 
\begin{tikzpicture}
\tikzstyle{every node}=[scale=3,font=\large]
\begin{axis}[
width=8in,
height=5in,
legend cell align={left},
legend style={
  legend columns=3, 
  fill opacity=0.8,
  draw opacity=1,
  text opacity=1,
  at={(0.5,1.18)},
  anchor=north,
  draw=white!80!black
},
ylabel style={text width=4.5cm, align=center},
tick align=outside,
tick pos=left,
x grid style={white!69.0196078431373!black, dashed},
xlabel={\Large Physical Resource Block},
xmajorgrids,
xmin=1, xmax=50,
xtick style={color=black},
xtick={10,20,...,50},
y grid style={white!69.0196078431373!black, dashed},
ylabel={\Large Received Interference Power [dBm]},
ymajorgrids,
ymin=-108, ymax=-95,
ytick style={color=black},
ytick={-108,-104,...,-95},
] 
\addplot [line width=3, red]
table {%
1	-104.4407693
2	-104.4407693
3	-104.4407693
4	-104.4407693
5	-107.1318253
6	-107.1318253
7	-107.2487522
8	-107.1318253
9	-106.6952397
10	-106.6952397
11	-106.2487522
12	-106.2487522
13	-106.2487522
14	-105.8008467
15	-105.4407693
16	-105.4407693
17	-105.2487522
18	-105.2487522
19	-104.1318253
20	-105.2487522
21	-105.2487522
22	-104.8008467
23	-104.5402967
24	-104.4407693
25	-104.2487522
26	-104.2487522
27	-104.2487522
28	-104.2487522
29	-103.5402967
30	-103.5402967
31	-103.5402967
32	-103.5402967
33	-103.4407693
34	-103.4407693
35	-103.4407693
36	-103.4407693
37	-103.4407693
38	-103.5402967
39	-103.8008467
40	-103.0629408
41	-103.0629408
42	-102.9736675
43	-103.0629408
44	-103.0629408
45	-102.9736675
46	-102.9736675
47	-104.4407693
48	-104.4407693
49	-104.4407693
50	-104.4407693
};
\addlegendentry{Intermod present}
 
\addplot [line width=3, black]
table {%
1	-101.345455
2	-101.345455
3	-101.345455
4	-101.345455
5	-101.906882
6	-101.8999757
7	-101.6596896
8	-100.8999757
9	-100.906882
10	-100.906882
11	-100.906882
12	-100.906882
13	-100.906882
14	-101.6596896
15	-101.6596896
16	-101.6596896
17	-100.906882
18	-100.906882
19	-100.8999757
20	-100.8999757
21	-100.8999757
22	-100.8912969
23	-100.7560127
24	-100.8999757
25	-101.6514774
26	-101.6514774
27	-101.6514774
28	-101.6514774
29	-101.6514774
30	-101.6514774
31	-101.6514774
32	-101.6514774
33	-101.641161
34	-100.628208
35	-100.628208
36	-100.641161
37	-100.6514774
38	-100.8912969
39	-100.8999757
40	-100.8999757
41	-101.6514774
42	-101.6514774
43	-101.641161
44	-101.641161
45	-101.4683616
46	-101.3302256
47	-101.345455
48	-101.345455
49	-101.345455
50	-101.345455
};
\addlegendentry{Not present}
\end{axis}
\end{tikzpicture}}\label{fig:ripprb}}
\hfil
\subfloat[]{
\resizebox{0.24\textwidth}{!}{\vspace*{-10em} 
\begin{tikzpicture}
\tikzstyle{every node}=[font=\large,scale=3]
\begin{axis}[
width=8in,
height=5in,
xmin=0, xmax=1,
xtick style={color=black},
xtick={0,0.2,...,1},
legend cell align={left},
legend style={
  fill opacity=0.8,
  draw opacity=1,
  text opacity=1,
  at={(0.97,0.03)},
  anchor=south east,
  draw=white!80!black
},
tick align=outside,
tick pos=left,
x grid style={white!65!black},
xlabel={False Positive Rate},
xmajorgrids,
xmin=0, xmax=1,
xtick style={color=black},
y grid style={white!65!black},
ylabel={True Positive Rate},
ymajorgrids,
ymin=0, ymax=1.05,
ytick={0,0.2,...,1},
ytick style={color=black}
]
\addplot [line width=3, blue]
table {%
0 0
0 0.75
0 0.75
0 1
0 1
0 1
0 0.75
0 0.75
0 1
0 1
0 1
0 0.75
0 0.75
0 1
0 0.75
0 0.75
0 1
0 1
0.5 1
1 1
1 1
};
\addlegendentry{\normalsize Area $= 1.0000$}
\addplot [line width=3, black, dashed, forget plot]
table {%
0 0
1 1
};
\end{axis}

\end{tikzpicture}}\label{fig:results_roc}}
\hfil
\subfloat[]{ 
\resizebox{0.18\textwidth}{!}{\vspace*{-10em} \begin{tikzpicture}[
box/.style={draw,rectangle,minimum size=2cm,text width=1.5cm,align=center}]
\matrix (conmat) [row sep=.1cm,column sep=.1cm] {
\node (tpos) [box,
    label={[rotate=90, xshift=0.7cm,left=0.25cm]left:\( \text{Present} \)},
    ] {6};
&
\node (fneg) [box
    ] {0};
\\
\node (fpos) [box,
    label={[rotate=90, xshift=0.95cm, left=0.25cm]left:\( \text{Not Present} \)},
    label=below:\(\text{Detected}\)] {0};
&
\node (tneg) [box,
    label=below:\(\text{Not Detected}\)
    ] {92};
\\
};
\end{tikzpicture}

}\label{fig:confusion_matrix}}
\hfil
\subfloat[]{ 
\resizebox{0.24\textwidth}{!}{
\begin{tikzpicture}
\tikzstyle{every node}=[font=\large,scale=3]
\begin{axis}[
width=8in,
height=5in,
legend cell align={left},
legend style={
  fill opacity=0.8,
  draw opacity=1,
  text opacity=1,
  at={(0.18,0.83)},
  anchor=south,
  draw=white!80!black,
  font=\tiny
},
tick align=outside,
tick pos=left,
x grid style={white!69.0196078431373!black, dashed},
xlabel={Number of PRBs},
xmajorgrids,
xmin=10, xmax=50,
xtick style={color=black},
xtick={10,20,...,50},
y grid style={white!69.0196078431373!black, dashed},
ylabel={Normalized run time},
ymajorgrids,
ymin=0, ymax=4,
ytick style={color=black},
ytick={0,1,...,4},
]
\addplot [black, mark=*, line width=2.5pt]
table {%
10    1.000000
20    1.910921
30    2.485139
40    2.858954
50    3.255480
};
\end{axis}

\end{tikzpicture}}\label{fig:runtime}}
\end{adjustwidth}
\caption{Simulation results: (a) RIP vs PRB for two cases: intermodulation interference is present and is not present (b) receiver operating characteristic plot (c) confusion matrix and (d) normalized run time vs number of PRBs.}
\end{figure*}
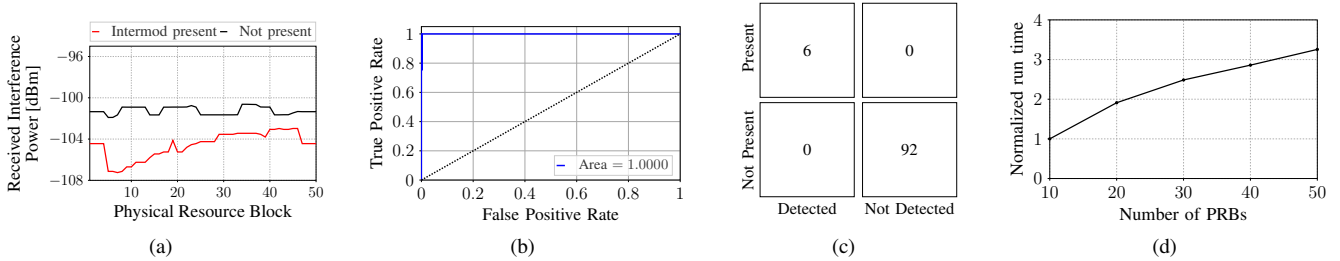%
\vspace*{-0.5em}
\subsection{Proposed} Let $\mathbf{R}$ be the user-plane RIPs measurements of a given BS collected periodically and sorted column-wise per PRB.  Thus, $\mathbf{R}\in\mathbb{R}^{\vert\mathcal{T}\vert\times N_\text{PRB}^{(u)}}$, where $\mathcal{T}$ is the set of measurements collection time window at a given periodicity (e.g., hourly) \cite{9500204}.

Let us formulate our proposed method as a detection problem.  Thus, a binary hypothesis testing is formulated where for a given BS, the null hypothesis $\mathcal{H}_0$ denotes absence of intermodulation interference while the alternative hypothesis $\mathcal{H}_1$ denotes the presence of intermodulation interference.  These instances of absence or presence are studied through the use of linear regression, which allows us to study if the RIP slopes with the PRBs.  We choose linear regression to avoid having to deal with training and test data.  Then, depending on the slope value $\beta_1$ and the score of fit ($R^2$) there are the following qualitative cases: 1) poor line fitting, 2) good line fitting and a small slope, and 3) good line fitting and a large slope. {It should be clear that case 3 represents $\mathcal{H}_1$.}

\textbf{Poor line fitting:} With $R^2 < \epsilon_\text{linear}$, the relationship between RIP and the PRBs is not linear enough to conclude the existence of intermodulation interference on the BS.  For example, there could be either 1) narrow-band interferers that impact only a select group of PRBs (which have low score of fit) or 2) peaks due to spectral leakage are in the receive band.

\textbf{Good line fitting:} With $R^2 \ge \epsilon_\text{linear}$, the relationship between RIP and the PRBs has enough linearity where the slope $\beta_1$ is meaningful.  In this case there are two potential outcomes: 1) $\vert \beta_1 \vert < \epsilon_\text{slope}$ which means that the slope is too low and therefore no intermodulation interference exists or 2)  $\vert \beta_1\vert \ge \epsilon_\text{slope}$ and therefore intermodulation interference exists.  In perspective, $R^2 \!=\! 0.81$ in the fitted regression line in Fig.~\ref{fig:freq}.

\textbf{Algorithm outline:} In Algorithm~\ref{alg:algorithm}, a row vector $\mathbf{r}$ from the matrix $\mathbf{R}$ is linearly regressed against the user-plane PRBs.  For example, in a BS with $N_\text{PRB}^{(u)} = 42$, an hourly measurement for this BS would be a row vector with $42$ entries.  The regression is used to suggest if intermodulation was detected for this measurement. This is repeated for all measurements in $\mathcal{T}\!$.

{\textbf{Parameter tuning:} The choice of $\epsilon_\text{linear}$ and $\epsilon_\text{slope}$ define the threshold at which we reject the null or alternative hypothesis. In order to find  detection optimal values of these parameters, a receiver operating characteristic (ROC) curve is constructed based on a grid search over the space values that these parameters can assume.  We select the values that correspond to the highest area under the ROC curve.  This effectively minimizes the errors of rejecting $\mathcal{H}_0$ when it is true (i.e., no intermodulation interference is present) and also minimizes the error of not rejecting $\mathcal{H}_0$ when it is false (i.e., intermodulation interference is likely present).  These are known as Type I and Type II errors, respectively.}

\textbf{Run-time complexity:} The run-time of the proposed algorithm for one BS given the two learning features is {linear in the number of PRBs in that BS}, as we show in Section~\ref{sec:simulation}.%

\textbf{Benefits:} Unlike the current methods, our proposed algorithm does not have limitations with respect to implementation time or impacting users in the service area of the serving BS or any of the neighboring BSs.  Furthermore, due to its constant run-time complexity, it can be run in real-time.%

\section{Simulation}\label{sec:simulation}%
\vspace*{-0.5em}
\subsection{Setup}
We use a mobile operator dataset comprising {two tables:} 1) hourly RIP {and RTP per receive branch 2) binary labels denoting whether intermodulation interference is detected through OCNG}.  This dataset has {$100$ rows representing measurement records of different BSs collected at different hours and days}.  Each BS has at least two frequency bands, each of $B = 10$ MHz and has {$N_\text{SC} = 12$}, $N_\text{PRB} = 50$, and $N_\text{PRB}^{(u)} = 42$. %
We drop any measurement record with an average RTP below $N_0 + 10\log B = -104$ dBm since these are unlikely to have experienced interference.
{To find the detection performance optimal value $\epsilon^\ast_\text{linear}$, we choose from a set of values that correspond to strong linear correlation $\{0.95, 0.9, 0.85, 0.8\}$.  Similarly, for $\epsilon^\ast_\text{slope}$, we choose from a set of values that correspond to high RTP $\{1.03, 1.05, 1.07, 1.09, 1.11\}$.  We normalize the linear regression features and enable the intercept.}



\subsection{Discussion}
{We start the discussion examining Fig.~\ref{fig:ripprb}.  This figure shows two RIP vs PRB plots.  The sloped line is suggestive of the presence of intermodulation, as motivated earlier.  The patterns superimposed on the lines are due to uplink PRB allocation (e.g., due to traffic) or other types of impairments.}

Next, we ask the following two questions: 1) how {well does the proposed} intermodulation interference {detection algorithm perform} in a realistic network? and {2) what is the impact of the number of PRBs on the run-time of the proposed algorithm?}

{To answer the first question, we compare the performance of our proposed algorithm against the labels.  The ROC curve in Fig.~\ref{fig:results_roc} shows perfect prediction compared to the OCNG detection method which means that the algorithm always rejects the false hypothesis and thus detects intermodulation when it is present. This shows that testing for a sloped line in the interference pattern (using linear regression) is sufficient for intermodulation interference detection purposes.  In perspective, there are $98$ records in the data (out of $100$) the average RTP of which is greater than $-104$ dBm.  Out of these, $6$ belong to BSs that have intermodulation interference while $92$ do not.  Our proposed algorithm correctly detects whether a measurement record represents intermodulation interference as shown in the confusion matrix in Fig.~\ref{fig:confusion_matrix}.}

Finally, answering the second question, Fig.~\ref{fig:runtime} shows that the proposed algorithm runs in $\mathcal{O}(N)$, where $N \coloneqq \vert \mathcal{P} \vert$, {making it suitable for advanced radio resource management procedures such as carrier aggregation.}

A follow-up question stemming from this discussion is: if the receive band were wide enough to cover not only the linear region due to spectral leakage, but also the main peak, or if other narrow-band interferers existed along the intermodulation. In such a case, how would the detection algorithm perform? The answer is this would no longer be a linear characterstic and the $R^2$ value of linear regression is likely to correspond to weak linear correlation.  Other supervised machine learning algorithms would then need to be used to account for the evident non-linearity, which may also have a non-linear run-time complexity.%

\section{Conclusion}\label{sec:conclusion}

In this paper, we proposed a novel method to detect the presence of intermodulation interference in an OFDM-based system without the need of test tones or noise simulators as specified in industry standards. We showed salient intermodulation spectral characteristics that are due to FFT.  This method uses machine learning in the form of linear regression of the received interference power as a function of the PRBs.  Based on this method, we constructed an algorithm that detects intermodulation interference in linear run time making it suitable for real-time radio resource management applications in O-RAN,  B5G, or 6G.  An interesting extension is when the receive frequency band experiences nonlinearities due to either spectral leakage or other narrow-band interferers.

\bibliographystyle{IEEEtran}
\bibliography{main.bib}

\end{document}